\documentclass{aastex}
\usepackage{geometry}
\usepackage{natbib}
\begin{document}
\title{Component-resolved Near-infrared Spectra of the (22) Kalliope System}
\shortauthors{Laver et al}
\author{Conor Laver$^1$, Imke de Pater$^1$, Franck Marchis$^{1,2}$, M\'at\'e \'Ad\'amkovics$^1$, and Michael H. Wong$^1$}
\affil{$^1$Department of Astronomy, 601 Campbell Hall, University of California, \\ Berkeley, CA 94720}
\affil{$^2$SETI Institute,  515 Whisman Rd, Mountain View, CA 94043}
\slugcomment{Pages: 20, Figures: 5, Tables: 1}

\pagebreak

\textbf{}\\
\noindent
\textbf{Proposed running head:}\\
Near IR spectra of the (22) Kalliope System
\\
\\
\\
\\
\noindent
\textbf{Please send editorial correspondance to:}\\
Conor Laver\\
Department of Astronomy\\
University of California\\
Berkeley, CA 94708, USA\\
\\
Email: conor@astro.berkeley.edu\\
Phone: +1 510 643 8591

\clearpage

\noindent

\textbf{Abstract:} We observed (22) Kalliope and its companion Linus
with the integral-field spectrograph OSIRIS, which is coupled to the
adaptive optics system at the W.M. Keck 2 telescope on March 25
2008. We present, for the first time, component-resolved spectra acquired simultaneously
in each of the Zbb (1-1.18$\mu$m), Jbb (1.18-1.42$\mu$m),
Hbb (1.47-1.80$\mu$m), and Kbb (1.97-2.38$\mu$m) bands.
The spectra of the two bodies are remarkably similar
and imply that both bodies were formed at the same time from the same
material; such as via incomplete re-accretion after a major impact on
the precursor body.


\clearpage

\section{Introduction}
The large main-belt asteroid (22) Kalliope, discovered in 1852, is
classified as a M-type asteroid \citep{tedesco89}.  Almost 150 years
after its discovery, it was found to have a small companion, Kalliope
I Linus (\citealp{merline01}, \citealp{margot01}), orbiting the
primary in a near-circular orbit with a semimajor axis of 1095 $\pm$
11 km in 3.596 $\pm$ 0.001 days \citep{marchis08}. For simplicity, in
this work we will refer to the primary of the (22) Kalliope  system as
``Kalliope" and its companion satellite as ``Linus".

Analysis of the orbit of Linus has been used to derive a mass of $8.1
\pm 0.2 \times 10^{18}$ kg for the primary \citep{marchis08}. In
addition to these direct observations of the binary system,
\citet{descamps08} used mutual eclipses during an equinox in 2007 and
a stellar occultation event observed in 2006 to constrain the sizes
and shapes of the two bodies, and hence the density of the
primary. Kalliope's equivalent radius is 83.1 $\pm$ 1.4 km and its
shape can be approximated by a triaxial ellipsoid with semimajor axes
of $117.5 \times 82 \times 62$ km. With this size and mass, Kalliope's
bulk density is 3.35 $\pm$ 0.33g/cm$^3$, significantly larger than
estimates for C-type or S-type binary asteroids derived so far
\citep{marchis08} and larger than previous estimates of Kalliope's
bulk density, which were based on a larger (IRAS-derived) size of the
object (\citealp{margot03,marchis03,marchis08}). Although \cite{descamps08}
derived a radius of 14 $\pm$ 1 km for Linus, this measurement was based 
on one eclipse detection taken at a particular viewing geometry. Since
the shape of Linus is unknown, we will adopt here the more conservative
estimate of 13$\pm$5.5 km based on multiple adaptive optics (AO) observations
\citep{marchis08}.

Since Kalliope is a M-type asteroid with almost featureless visible
and near-infrared spectra and a high albedo \citep{marchislpi}, it is
difficult to assess its meteorite analog and derive its
macro-porosity. An upper limit of 50--60\% for the porosity can be
obtained by assuming a pure Ni-Fe asteroid \citep{britt02}. Based upon
the theoretical work by \cite{wilson99} with regard to gravitational
re-accretion of bodies after a complete disruption, \cite{descamps08}
adopt a porosity of 20--40\% for Kalliope, i.e., they assume the body
is at least heavily fractured or perhaps a rubble-pile. Such a
porosity implies a grain density between 4 and 6 g/cm$^3$, suggestive
of a mixture of Ni-Fe alloys and silicates, consistent with near- and
mid-IR spectroscopy \citep{marchislpi}.  Such a heavily fractured
primary, combined with the retrograde nature of the secondary's orbit
suggest that the system may have formed from a large impact on a
proto-Kalliope \citep{durda04}.

With the increasing number of known multiple systems in all
populations of small solar system bodies, one can begin to
statistically analyze the distribution of mass ratios and orbital
characteristics, which provide constraints on the formation of such
systems. This is an important step towards developing an accurate
picture of the environment during the solar system's formative
period. It is particularly helpful to determine whether these
asteroidal satellites formed simultaneously with the
primary, through catastrophic impacts, or via
capture. Component-resolved color ratio measurements are used to 
determine if an asteroid and its companion have the
same surface composition. For example, \cite{marchis06} report
identical color ratios for the double trojan asteroid (617)
Patroclus-Menoetius, which suggests a similar surface
composition. However, without detailed spectroscopic measurements, such
inferences remain speculative. Moreover, it is difficult to
make accurate flux measurements if the secondary is
faint ($3<\Delta m<7$) and close (0.3-0.7") to the primary. 

A spectroscopic comparison of a primary and its satellite in binary
systems such as Kalliope-Linus, may help to distinguish between the
various formation scenarios. Similar spectra would indicate that the
formation of both bodies was likely in situ or through a catastrophic
collision, whereas measurably different spectra would point towards a
foreign body capture. In order to constrain the origin of the
binary asteroid system (22) Kalliope, we use the field-integral
spectrograph OSIRIS (OH-Suppressing Infra-Red Imaging Spectrograph), on
the W.M.  Keck II 10-m telescope, to combine the high angular
resolution provided by the telescope's adaptive optics system with
spectroscopy. The observations and the data analysis are presented in
Sections 2 and 3 respectively, with a discussion of the results in
Section 4.

\section{Observations}\label{sec:obs}

We observed (22) Kalliope and its satellite Linus on 25 March 2008
between 07:33 and 08:13 UT using OSIRIS at the W.M. Keck observatory
in Hawaii. OSIRIS is equipped with a 2048 x 2048 pixel Rockwell
Hawaii-2 detector \citep{krabbe06} and covers a wavelength range from
0.9 $\mu$m to 2.4 $\mu$m. In each of four broadband filters (Zbb, Jbb,
Hbb and Kbb), OSIRIS records data cubes with 2 spatial and 1 spectral
dimensions. We present observations in each of these broadband
filters using the 0.02" platescale, which provides a field-of-view
of $0.32^{\prime\prime} \times 1.28^{\prime\prime}$ (Table
\ref{tab:obs}). The spectral resolution was $R = \frac{\lambda}{\Delta
  \lambda} \approx 3800$. The angular resolution depends on the
brightness of the target (V-mag for wavefront sensing) and the
atmospheric conditions. We estimate a spatial resolution of 0.06"
from observations of nearby PSF (point spread function) stars.

\begin{figure}[!ht]

\includegraphics{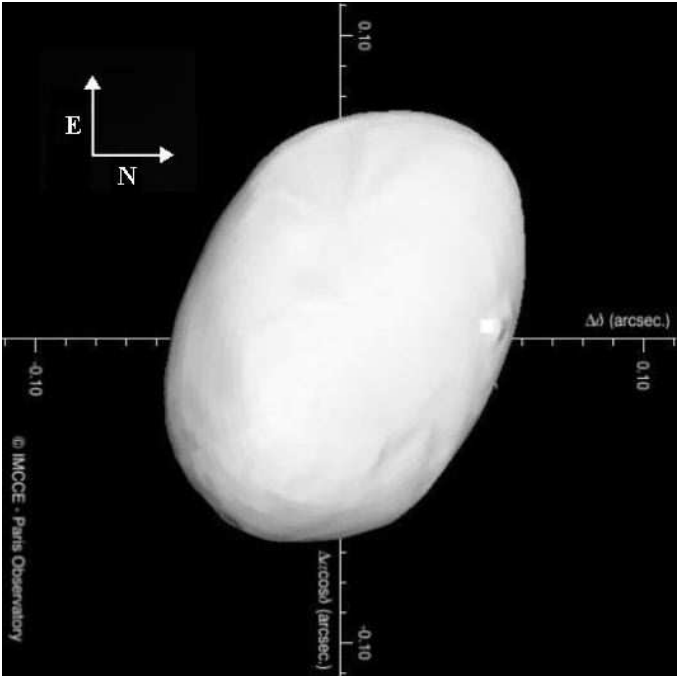}

\caption{\label{fig:model}Generated image showing Kalliope's primary  appearance at 07:47UT on March 25 2008 according to the \cite{descamps08} 3D-shape model. The image has been rotated to match the observed position angle in Fig. \ref{fig:im} (Generated using AsteROT, P. Descamps IMCCE).}

\end{figure}

\begin{figure}[!ht]

\includegraphics[scale=0.8]{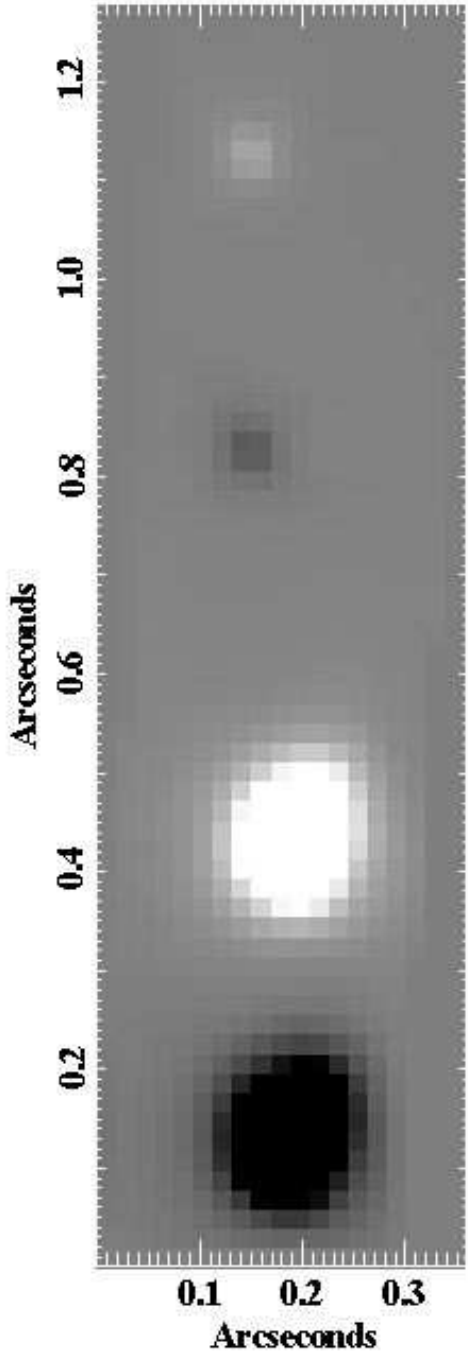}

\caption{\label{fig:im}Wavelength averaged Hbb datacube subtraction showing Kalliope (bottom) and its companion (top), Linus, in two nod positions (positive and negative after subtraction). The figure has a normalized intensity on a linear scale.}

\end{figure}

At the time of the observations, Kalliope was at a geocentric distance
$\Delta= 2.083$ AU, heliocentric distance $r =3.048$ AU, and was
observed almost pole-on according to the pole-solution and shape model
derived in Descamps et al. (2008). This 3D-shape model predicts a
projected equivalent diameter of 194$\pm$2km throughout the 45 min of
observations. A sample model image of Kalliope at 07:47UT is shown in
Fig. \ref{fig:model}. The angular resolution of 0.06" leads to a
spatial resolution of $\sim$91 km at Kalliope's
distance. Consequently, Kalliope's primary is resolved in our image
data cubes, while Linus is a point source. The $\sim$1100 km
semi-major axis of the satellite's orbit corresponds
to $\sim$0.72" for a perfectly pole-on orientation. The binary system
is therefore well resolved in our data (see Fig. \ref{fig:im}).

The data are reduced with the OSIRIS pipeline \citep{larkin06}. This
pipeline corrects for instrumental effects before rectifying and
calibrating the datacubes in preparation for analysis. Nodded postions
of the target are subtracted for correction of the sky background
contamination. Sample images are shown in Fig. \ref{fig:im}, where the
datacube is averaged over all wavelengths in the filter. Inverted 
images of the targets arise from the nodded sky subtraction.
The extraction discussed below is performed for each target
individually and the inverted spectra are averaged with the positive 
spectra to provide one final spectrum per target. 

At each wavelength, aperture photometry was performed on both science
targets and corrected for any residual background. This provides a raw
spectrum of Kalliope and Linus. The proximity of the secondary object
and the observing method of nodding the target up and down the slit
for background subtraction, both possibly lead to some contamination
of the secondary's spectra. This is exacerbated in the Zbb filter
where the primary halo is brightest because the AO performance
(wavefront correction) is not as good as at the longer wavelengths; in addition, the primary is much brighter because the
solar flux is higher. More accurate flux measurements can
be obtained by nodding completely off the object for sky
calibration (but this decreases the total on-source integration time).

We corrected both primary and secondary object spectra for telluric
absorption in Earth's atmosphere, and photometrically calibrated them
in a manner similar to that described by \cite{laver08}, using 
observed spectra of the stars HD77281 (type A2) and HD111133 (type
A0), both 2MASS catalog stars (see Table
\ref{tab:obs}). The telluric correction was determined by
comparing the spectra of the stars to the calibrated spectra from
stellar atmosphere models. These spectra were created using the 
Kurucz stellar
models\footnote{ref. http://kurucz.harvard.edu/stars}, which were
adjusted for reddening using a Fitzpatrick reddening curve
\citep{fitz99}. The model spectra were subsequently scaled to match
the 2MASS J,H,K magnitudes for the target stars \citep{cutri03}. The
wavelength-dependent factor needed to scale the observations of the
stars to match the models is used to calibrate the target
spectra. Although we use both stars to estimate the telluric
correction, differences in time and airmass result in imperfect
corrections near the edges of some of the filters where
water and CO$_{2}$ bands dominate.

As Kalliope and Linus are seen in reflected solar light, we corrected
the spectra for solar features by dividing them by a normalized solar
spectrum, and a solar blackbody curve. We used the high resolution
solar spectrum from the ISAAC
website\footnote{ref. http://www.eso.org/sci/facilities/paranal/instruments/isaac/tools/},
which was convolved to the spectral resolution of the OSIRIS
instrument. Rather than calibrating the data by observing G-type
stars, we use a combination of library solar spectra with
A-type stars for two reasons: A-star spectra have fewer absorption 
lines than G stars, and because other G stars are not perfect solar 
analogs. In general these steps are
sufficient to produce high-sensitivity spectra; however, slight
imperfections in the data rectification matrices of the Jbb and Zbb
filters occasionally result in corrupted data, manifested as
single aberrant pixels in the cube. Where applicable these outliers
have been cleaned up during the data processing.

\begin{figure}[!ht]

\includegraphics[scale=0.8]{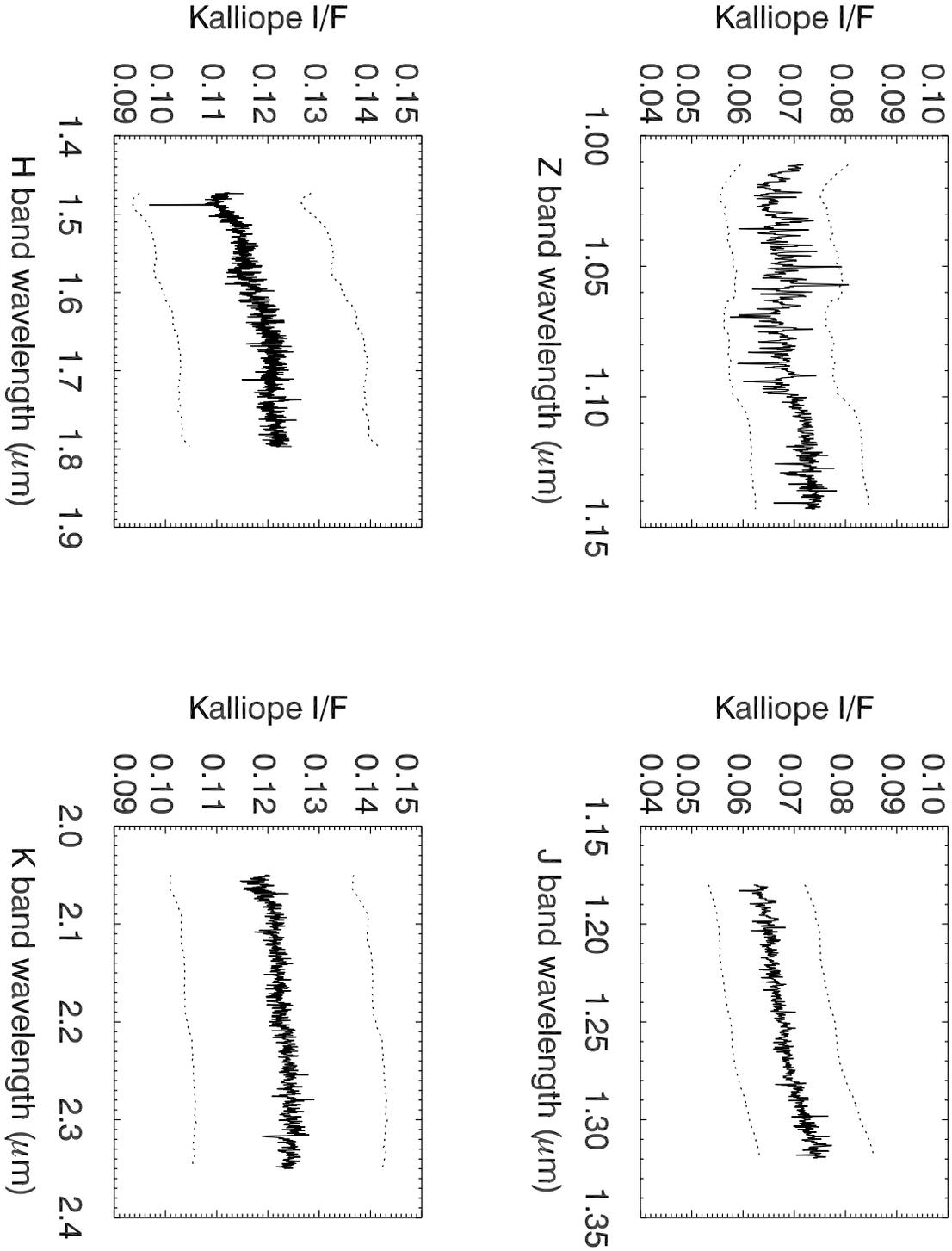}

\caption{\label{fig:kal}Zbb, Jbb, Hbb and Kbb spectra of Kalliope shown in $I/F$, using the apparent diameter of 194km. Dashed lines show the effect of the photometric calibration uncertainty, which is wavelength independent. }

\end{figure}

\begin{figure}[!ht]

\includegraphics[scale=0.8]{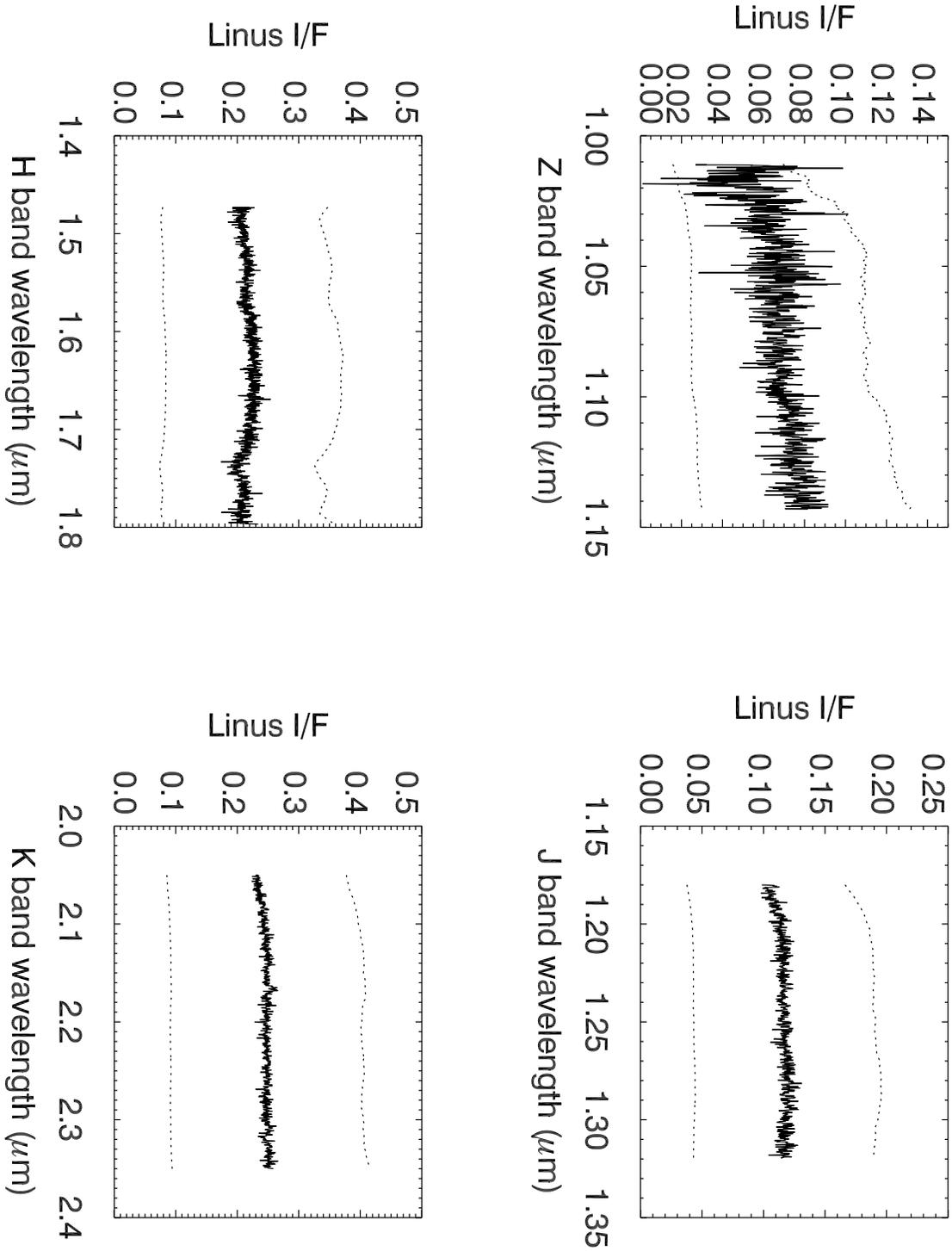}

\caption{\label{fig:lin}Zbb, Jbb, Hbb and Kbb spectra of Linus shown in $I/F$, using a diameter of 26 $\pm$11km. Dashed lines show the effect of the size uncertainty, combined with the photometric calibration errors, which is wavelength independent.}

\end{figure}

The units on the spectra were converted to the dimensionless parameter
$I/F$, where $I$ is the reflected intensity and
$\pi F$ the solar flux density at Kalliope's distance \citep{hammel89}:
\begin{equation}
{I \over F} = {r^2 \over \Omega} {F_p \over F_\odot},\label{eq1}
\end{equation}

\noindent with $r$ Kalliope's heliocentric distance, $\pi F_\odot$ the
Sun's flux density at Earth's orbit, $F_p$ the observed flux density
from the object, and $\Omega$ is the solid angle (in steradians)
subtended by the body ($\pi R^2/\Delta^2$). By this definition, $I/F=1$
for uniformly diffuse scattering from a (Lambert) surface when viewed
at normal incidence, and is equal to the geometric albedo if the
object is observed at a solar phase angle of $0^\circ$. To determine
$\Omega$, we used Kalliope's effective size as derived from
\cite{descamps08}'s models. The spin rate and shape of Linus are not
well known and thus the $I/F$ of Linus inherits the large error bars of
the estimated diameter (26 $\pm$ 11km). We also note that the
photometry of OSIRIS data at the 20 mas platescale is not perfect, due
to a loss of flux into the PSF halo. Although to first approximation
our calibration procedure has taken this into account, we adopt a 15\%
uncertainty in our absolute flux calibration, based on previous OSIRIS
observations \citep{laver08}. Our results are shown in
Figs. \ref{fig:kal} and \ref{fig:lin} for Kalliope and Linus,
respectively.

\section{Analysis}

Both bodies are slightly reddish, with an $I/F$ of 0.05-0.15, which is
typical for M-type asteroids \citep{gaffey89}.  Kalliope's size is
based on the \cite{descamps08} model, which predicts a nearly pole-on
configuration during our observations, with no discernible change in
the projected equivalent diameter, even though the observations span
almost 1/4 of Kalliope's spin period (which is 4.18 hrs). The shape
and spin of the secondary are unknown, yet it is reasonable to
assume that Linus is tidally locked (such that the spin period equals
the orbital period of $\approx$ 3.6 days), given its proximity to the
primary  ($a \sim$ 12 $\times$ R$_{primary}$) \citep{marchis08} and its
low orbital inclination relative to Kalliope's equatorial plane
($\sim 0^\circ$).

There are small differences in the absolute $I/F$ between filters,
which may be larger than the red slope expected for a M-type
asteroid. Such differences may stem from either errors in the above
assumptions about Linus, the Kalliope model, or simply the photometric
uncertainties in some of the filters. In addition, the calibration of
the slopes in spectral datacubes from OSIRIS may be somewhat
problematic. With a small field of view (0.32"x1.28") it is easy to
encounter difficulties with the wings of a PSF falling off the edges
of the detector. While this makes photometric calibration more
challenging, it can also affect the overall slope of the spectrum as
PSFs broaden at longer wavelengths, and hence more light can be 'lost'
at longer wavelengths, i.e., the spectra may in fact be slightly more
red than shown. However, while it is difficult to quantify such
effects, these have been largely corrected for in our calibration
scheme as we used entire spectra for photometric calibration (and our
science targets are only slightly resolved).

The ratio of the spectra is, however, much more stable as the slopes
of the spectra of each target are affected in a similar way. The
ratios of the spectra of Kalliope and Linus are plotted in Fig
\ref{fig:comp}; these reveal a remarkably flat slope.  The flux ratio
is $\sim$30 in all three J, H, and K bands, which equates to a
difference in magnitude of 3.7 $\pm$0.1. The Zbb spectrum of Linus has
a lower intensity (relative to longer wavelengths) and a much noisier
spectrum. This is caused by the brighter halo of the primary at Zbb
wavelengths due to the poor correction provided by the AO system at
shorter wavelengths, which increases the background noise in the
vicinity of the secondary. The flux ratios, as indicated in Table
\ref{tab:obs} for each filter, depend upon the relative sizes and
albedos of the primary and secondary. Given the uncertainties in the
size of Linus, however, it is difficult to extract information about
the relative albedos of the bodies. Assuming similar albedos for the
two bodies, as one might expect based on the similarity of their
spectral shapes, we derive an effective diameter of 35$\pm$2km for
Linus on the night of the observations, which is consistent with the
size estimate of 26$\pm$11km.

We see an apparent absorption line at 1.49 $\mu$m in Kalliope's
spectrum (Fig. \ref{fig:kal}). This line occurs at the same location
as a moderately strong solar absorption line (Mg I line at 1.48817
microns) and the narrowness of the feature leads suggests that this
line is not intrinsic to the asteroid but is an artifact of the
reduction process. However, all other solar lines of similar strength
are correctly removed, so the presence of this line remains
unexplained. The remaining apparent features have all been attributed
to incomplete telluric correction due to the variation in the
atmosphere between the science target and calibration stars.

\section{Discussion}

\begin{figure}[!ht]

\includegraphics[scale=0.7]{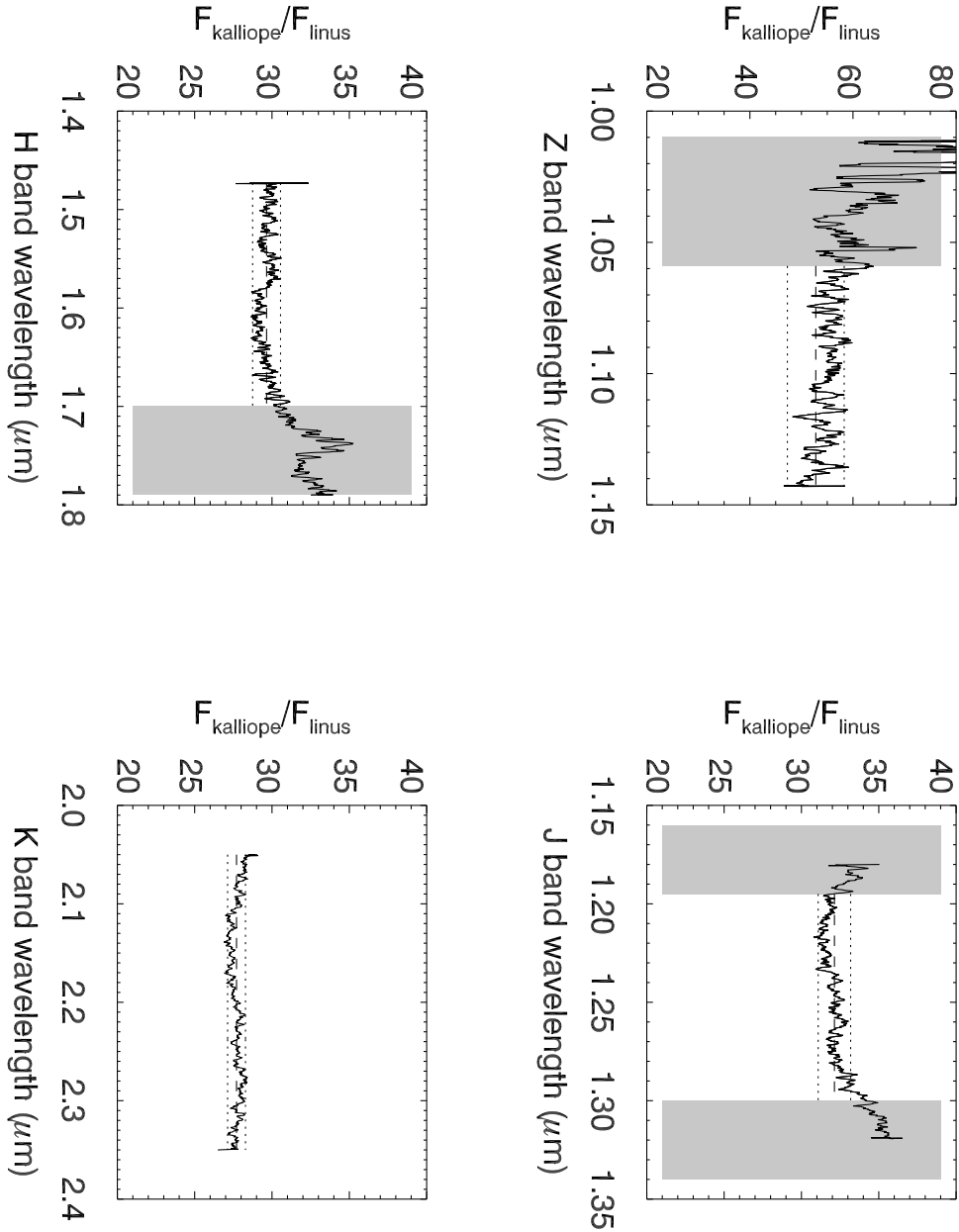}

\caption{\label{fig:comp}Zbb, Jbb, Hbb and Kbb band comparisons of the primary and secondary spectra. The data have been smoothed using a 10 pixel running boxcar. The grey areas indicate with strong telluric absorption which is difficult to fully correct. The dashed lines indicate the average flux ratio(with 1$\sigma$ errors) in the regions of the spectra least affected by the telluric correction. }

\end{figure}

The spectra in Fig. \ref{fig:kal} confirm previous findings that
Kalliope is a M type asteroid, devoid of silicate
features. Potentially interesting hydrated features in the 3 $\mu$m
wavelength range, which may be present on Kalliope \citep{rivkin00},
are outside the wavelength range of the OSIRIS instrument. Linus
displays a very similar spectrum as shown in Figs. \ref{fig:lin} and
\ref{fig:comp}. This implies that the surface material on both bodies
is similar in composition and may have undergone similar degrees
of space weathering.

Using the orbital characteristics of Linus and Kalliope,
\cite{descamps08} showed that (22) Kalliope may be one of the oldest
binary asteroid systems, with an estimated age between 1 and 3
Gyrs. Hence a study of this system provides a window into the
conditions in the early solar system. It is apparent from
Fig. \ref{fig:comp} that the spectra of Kalliope and Linus are
essentially the same, implying that both bodies formed at the same
time from the same material. One explanation is that the
Kalliope-Linus system formed from the same parent body. A
proto-Kalliope could have been disrupted in a catastrophic impact, and
re-accretion of material led to the formation of Kalliope and its
satellite \citep{margot03, marchis03,marchis08}. Due to Kalliope's
weak self gravity, this re-accretion naturally leads to an irregular,
heavily fractured or rubble-pile body \citep{durda04}.

\cite{descamps08} argued that an oblique impact on an already
fractured or porous proto-Kalliope could also lead to the formation of
a binary system via fission. If this were the case, one might expect
space weathering effects to be different on the primary and
secondary surfaces, unlike the catastrophic impact theory. There is no
clear measurement of the space weathering effect on M-type asteroids,
mostly because we do not know their composition and we do not know any
young collisional families from this taxonomic
class. \cite{vernazza2009} studied the reddening of the spectra for
young S-type collisional families. They showed that after a fast
reddening due to ion implantation ($\sim$1 Myr), the spectra of
asteroids continue to change (`mature') at a slower pace by micro-meteorite
impacts (up to 2,500 Myr). The bulk density and mid-infrared
emissivity spectrum of (22) Kalliope \citep{marchislpi} suggest that
the asteroid is composed of a mixture of enstatite, iron and also
olivine which is 'weathered' in S-type asteroids. If this is true,
we can conclude from the striking similarities between the
near-infrared spectra of the two components of (22) Kalliope, that the
satellite and the primary likely formed at approximately the same time
from the same parent body, and that the ejecta scenario is highly
unlikely.

\section{Conclusion}

This paper presents the first component-resolved spectra of (22)
Kalliope and its satellite Linus, obtained with the field-integral
spectrograph OSIRIS at the W.M. 10-m Keck telescope. The data reveal
remarkably similar M-type spectra for the two objects in the Zbb, Jbb,
Hbb and Kbb infrared filters. This similarity between the spectra
implies a common origin for the two bodies, i.e., they must have
formed at the same time from the same materials, most likely via a
catastrophic disruption of a proto-Kalliope. The relative intensity
of the two objects differs by a factor of $\sim$30, or 3.7 magnitudes
which is consistent with the size estimates of the primary
(D$_{app}$=194 km) at the time of the observation and the satellite
(D=26$\pm$11 km). If similar albedos are assumed this relative
intensity implies a effective diameter of 35$\pm$2km for Linus.

Our study shows the value of using an integral field spectrometer
combined with an adaptive optics system, to simultaneously record
spectra of several objects in multiple asteroid systems. A spectral
survey of multiple asteroid systems will reveal if the observed
similarity of the spectra of Kalliope-Linus is typical for binary
asteroids, or if there is a large variety amongst them. Such surveys
are therefore important to constrain theories on the formation and
evolution of multiple asteroid systems, and hence provide information
on the environment of our early solar system.

\section{Acknowledgments}

This work was supported in part by the National Science Foundation Science and Technology Center for Adaptive Optics, managed by the University of California at Santa Cruz under cooperative agreement AST 98-76783. FM acknowledges additional support from NASA grant NNX07AP70G and NSF grant AAG-0807468. CL thanks the American Museum of Natural History for their support and the use of their facilities. The data were obtained with the W.M. Keck Observatory, which is operated by the California Institute of Technology, the University of California, Berkeley and the National Aeronautics and Space Administration. The observatory was made possible by the generous financial support of the W.M. Keck Foundation. The authors extend special thanks to those of Hawaiian ancestry on whose sacred mountain we are privileged to be guests. Without their generous hospitality, none of the observations presented would have been possible.

\bibliographystyle{apalike}

\begin{deluxetable}{ccccccc}

\tablecaption{\label{tab:obs}Observations of the (22) Kalliope system on UT 25 March 2008. The flux ratios as observed for Kalliope/Linus are indicated$^1$.}

\tablehead{\colhead{Date} & \colhead{Start} & \colhead{Object} & \colhead{Filter/Scale} & \colhead{Int Time}  & \colhead{Airmass} & \colhead{Flux ratio}  \\
 \colhead{(UT)}  &  \colhead{(UT)} & \colhead{} & \colhead{} & \colhead{(coadds x s)} & \colhead{} & \colhead{Kalliope/Linus}} 

\startdata

Mar 25 2008 & 06:26& HD77281 & Kbb/0.02" & 1 x 30 & 1.10 & \\

Mar 25 2008 & 06:33& HD77281 & Hbb/0.02" & 1 x 30 & 1.09 & \\

Mar 25 2008 & 07:33& Kalliope & Hbb/0.02" & 1 x 300 & 1.54 & 30$\pm$1\\

Mar 25 2008  & 07:47& Kalliope & Kbb/0.02" & 1 x 300  &1.46 & 28$\pm$1\\

Mar 25 2008  & 08:00& Kalliope & Jbb/0.02" & 1 x 300  & 1.38 & 32$\pm$1\\

Mar 25 2008  & 08:13& Kalliope & Zbb/0.02" & 1 x 300  & 1.32 & 53$\pm$1\\

Mar 25 2008 & 08:32 & HD111133 & Kbb/0.02" & 1 x 30 & 1.26& \\

Mar 25 2008 & 08:36 & HD111133 & Hbb/0.02" & 1 x 30 & 1.24 & \\
Mar 25 2008 & 08:41 & HD111133 & Jbb/0.02" & 1 x 30 & 1.22 & \\
Mar 25 2008 & 08:46 & HD111133 & Zbb/0.02" & 1 x 30 & 1.21 & \\
\enddata


\end{deluxetable}

\clearpage 

\clearpage

\end{document}